\documentclass[preprintnumbers,jcp,showpacs,superscriptaddress]{revtex4-1}

\usepackage{color}
\usepackage{palatino}
\usepackage{dcolumn}
\usepackage[latin1]{inputenc}
\usepackage{epsf}
\usepackage{amsmath,amssymb,amscd,amsthm}
\usepackage{latexsym}
\usepackage{calc}
\usepackage[usenames,dvipsnames]{xcolor}
\usepackage{epsfig}
\usepackage[pdftex, pdfstartview = {FitH}]{hyperref}
\usepackage{graphicx}
\usepackage{amsfonts}
\usepackage{appendix}
\usepackage{lastpage}
\usepackage{fancyhdr}
\usepackage{enumerate}
\usepackage{ulem}

\usepackage{bm}

\def\bA{{\bf A}}
\newcommand{\dulR}{\mathbf R}
\newcommand{\dulr}{\mathbf r}
\newcommand{\bR}{\mathbf R}
\newcommand{\br}{\mathbf r}

\begin{document}
\title{The adiabatic limit of the exact factorization of the electron-nuclear wave function}

\author{F. G. Eich}
\affiliation{Max Planck Institute for the Structure and Dynamics of Matter, Luruper Chaussee 149, D-22761 Hamburg, Germany}
\author{Federica Agostini}
\email{agostini@mpi-halle.mpg.de}
\affiliation{Max Planck Institute of Microstructure Physics, Weinberg 2, D-06120 Halle, Germany}

\date{\today}
\pacs{}
\begin{abstract}
We propose a procedure to analyze the relation between the exact factorization of the electron-nuclear wave function and the Born-Oppenheimer approximation. We define the \textsl{adiabatic limit} as the limit of infinite nuclear mass. To this end, we introduce a unit system that singles out the dependence on the electron-nuclear mass ratio of each term appearing in the equations of the exact factorization. We observe how non-adiabatic effects induced by the coupling to the nuclear motion affect electronic properties and we analyze the leading term, connecting it to the classical nuclear momentum. Its dependence on the mass ratio is tested numerically on a model of proton-coupled electron transfer in different non-adiabatic regimes.
\end{abstract}
\maketitle 

\section{Introduction}\label{sec: intro}
A landmark in the description of the coupled motion of electrons and nuclei in molecular systems was the  seminal paper by Born and Oppenheimer~\cite{BO}. In the so-called Born-Oppenheimer (BO) method the time-independent Schr\"odinger equation is solved by expanding in orders of the electron-nuclear mass ratio. The lowest order equation corresponds to solving the electronic problem for fixed nuclei, while higher orders allow to consider nuclear vibrations, rotations and their coupling to the electronic degrees of freedom. What is nowadays usually referred to as the BO approximation (BOA) is the combined solution of (i) the electronic eigenvalue problem at fixed nuclear positions, i.e. the lowest order of the BO time-independent expansion, and (ii) the nuclear time-(in)dependent Schr\"odinger equation, where the potential is provided by a single eigenvalue of the electronic Hamiltonian, the so-called BO Hamiltonian, at each nuclear position~\footnote{The diagonal correction to the Born-Oppenheimer eigenvalue, due to the nuclear kinetic energy operator, will not be considered part of the Born-Oppenheimer approximation.}. In the time-dependent version of the BOA, the electronic and nuclear equations are derived from the full Schr\"odinger equation as consequence of an approximation, based on the adiabatic hypothesis: it assumes that the electrons react instantaneously to the motion of the nuclei. Therefore, they ``statically'' occupy an eigenstate corresponding to a given configuration of the nuclei.

The BOA is the cornerstone of molecular dynamics, because it allows to simulate a variety of complex molecular time-dependent phenomena efficiently and accurately. However, even in those situations where it conveys an overall correct description of the problem, a treatment based on the BOA is not free of inconsistencies. A few examples have been reported in the literature: the electronic mass is not properly accounted for in the dynamics of the nuclei~\cite{Kutzelnigg_MP1997, Subotnik_JPCL2012, Scherrer_TBS2016}, which can lead to inaccuracies when calculating vibrational and rotational excitation energies of light molecules~\cite{Moss_MP1997, Schwenke_JPCA2001}; in the absence of degeneracies and external magnetic fields, the electronic state computed within the BOA is real. As a consequence, the electronic current density vanishes and the continuity equation is violated~\cite{Barth_CPL2009, Paulus_JPCB2011, Schild_CP2010, Patchkovskii_JCP2012, Schild_JPCA2016, Bredtmann_PCCP2015}. For similar reasons, the vibrational circular dichroism computed within the BOA requires corrections~\cite{Scherrer_JCTC2013, Scherrer_JCP2015, Nafie_JPCA1997, Nafie_JCP1983} to yield agreement with experimental measurements. Various approaches~\cite{Nafie_JCP1983, Schild_JPCA2016, Patchkovskii_JCP2012, Moss_MP1997, Scherrer_TBS2016, Scherrer_JCP2015, Stephens_JPC1985, okuyama09, pohl15} have been proposed to cure these inconsistencies perturbatively. In this paper we further develop this idea by focusing on the following questions: Is there a procedure to systematically identify corrections to the BOA? Can a ``small'' parameter be identified that justifies that treatment of such corrections as perturbation? What physical quantities are affected by the corrections?

The starting point of the analysis is the exact factorization (XF) of the electron-nuclear wave function~\cite{Gross_PRL2010,*Gross_JCP2012}. In this framework, the solution of the time-dependent Schr\"odinger equation (TDSE) is written as a single product of a nuclear wave function and an electronic factor with parametric dependence on the nuclear configuration. The full TDSE splits into two coupled equations for the nuclear and for the electronic components. Such factorization of the molecular wave function is exact, thus it incorporates non-adiabatic effects arising from the coupling between electronic and nuclear motion. The form of the electron-nuclear wave function in the XF framework looks deceptively similar to the electron-nuclear wave function in the BOA. It is therefore natural to ask under which conditions the BOA is recovered from the XF. We will show that BOA emerges from the XF in the \textsl{adiabatic limit}~\cite{Hagedorn_AM1986, Hagedorn_AIHP1987, Hagedorn_PSPM2007, Lee_CPL1996}, i.e., when the electron-nuclear mass ratio tends to zero. Strong indications that this is indeed the correct limiting procedure to recover the BOA from the XF have been reported in a recent numerical study~\cite{Min_PRL2014}, where it was shown that characteristic features of the BOA, related to molecular geometric phases, only appear in the limit of infinite nuclear masses in the XF, except if one chooses current-carrying eigenstates in systems with degeneracies~\cite{Requist_PRA2016}.

In the framework provided by the XF, the nuclear equation is a TDSE, where a time-dependent vector potential (TDVP) and a time-dependent scalar potential, also referred to as time-dependent potential energy surface (TDPES), mediate the coupling to the electrons. The electronic equation is a less standard evolution equation, where an electron-nuclear coupling operator (ENCO), which acts on the parametric dependence of the electronic factor, is responsible for inducing non-adiabatic transitions. The nuclear equation describes how the nuclei evolve, while the electronic equation describes how the electrons follow the nuclear motion. This is consequence of choosing the product form of the full wave function as a \textsl{proper} nuclear wave function and a \textsl{conditional} electronic wave function, similarly to what is done within the BOA. This way of decomposing the electron-nuclear problem provides the same perspective as the BOA.
In the electronic equation, when the ENCO vanishes, the explicit coupling to the nuclear motion vanishes as well. Therefore, the exact electronic equation is expected to yield the eigenvalue problem associated to the BO Hamiltonian at fixed nuclear configuration.
We will show that this situation is formally recovered in the adiabatic limit.

The paper is organized as follows. In Section~\ref{sec: theory}, we briefly recall the XF and compare it to the BOA. In Section~\ref{sec: units}, we introduce the unit system used to analyze the dependence on the electron-nuclear mass ratio of the equations of the XF. We analyze this dependence in the electronic equation in Section~\ref{sec: electronic equation} and in the nuclear equation in Section~\ref{sec: nuclear equation}. 
The theoretical discussion is supported by numerical results, presented in Section~\ref{sec: results}. Conclusions are stated in Section~\ref{sec: conclusions}.

\section{Exact factorization of the electron-nuclear wave function}\label{sec: theory}
The XF has been presented~\cite{Gross_PRL2010,Gross_JCP2012} and extensively discussed~\cite{Gross_PRL2013, Gross_MP2013, *Gross_JCP2015, Gross_EPL2014,*Gross_JCP2014, Suzuki_PRA2014, *Suzuki_PCCP2015} in previous work. Therefore, we only introduce here the basic formalism and we refer to the above references for a detailed presentation.

A system of interacting electrons and nuclei, in the absence of  
external fields, is described by the Hamiltonian 
\begin{align}\label{eqn: hamiltonian}
\hat H = \hat T_n + \hat H_{\mathrm{BO}},
\end{align}
where $\hat T_n$ is the nuclear kinetic energy and $\hat H_{\mathrm{BO}}$ is the BO Hamiltonian, containing the electronic kinetic energy and all interactions. The evolution of the electron-nuclear wave function, $\Psi$, is described by the TDSE $\hat H \Psi = i\hbar\partial_t\Psi$. In the XF framework the solution of this TDSE is written as
\begin{align}\label{eqn: factorization}
\Psi(\dulr,\dulR,t)=\Phi_{\dulR}(\dulr,t)\chi(\dulR,t),
\end{align}
where $\chi(\dulR,t)$ is the nuclear wave function and $\Phi_{\dulR}(\dulr,t)$ is an electronic factor parametrically depending on the nuclear configuration $\dulR$. Here, the symbols $\dulr,\dulR$ have been used to collectively indicate the positions of $N_{el}$ electrons and $N_n$ nuclei, respectively. $\Phi_{\dulR}(\dulr,t)$ satisfies the partial normalization condition
\begin{align}
\int d\dulr |\Phi_{\dulR}(\dulr,t)|^2=1\quad\forall\dulR,t,
\end{align}
which makes the factorization in Eq.~(\ref{eqn: factorization}) unique up to a gauge-like phase transformation. Starting from the TDSE for the full wave function and using Frenkel's action principle~\cite{frenkel}, where the partial normalization condition is imposed by means of Lagrange multipliers~\cite{Alonso_JCP2013,*Gross_JCP2013}, the evolution equations for $\Phi_{\dulR}(\dulr,t)$ and $\chi(\dulR,t)$ are
\begin{subequations} \label{eqn: exact eqn}
\begin{align}
\left[\hat H_{\mathrm{BO}}+\hat U_{\mathrm{en}}\left[\Phi_{\dulR},\chi\right]-\epsilon(\dulR,t)\right] \Phi_{\dulR}(\dulr,t)&=i\hbar\partial_t\Phi_{\dulR}(\dulr,t)\label{eqn: exact electronic eqn}\\
\left[\sum_{\nu=1}^{N_n}\frac{\left[-i\hbar\nabla_\nu+\bA_\nu(\dulR,t)\right]^2}{2M_\nu}+\epsilon(\dulR,t)\right]\chi(\dulR,t)&=i\hbar\partial_t\chi(\dulR,t).\label{eqn: exact nuclear eqn}
\end{align}
\end{subequations}
Note that $\chi(\dulR,t)$ fulfills a standard TDSE. The evolution of the electronic factor $\Phi_{\dulR}(\dulr,t)$, however, is governed by a less common equation. In fact, the difference to a TDSE is the presence of the so-called electron-nuclear coupling operator (ENCO),
\begin{align}\label{eqn: enco}
\hat U_{\mathrm{en}}\left[\Phi_{\dulR},\chi\right]=&\sum_{\nu=1}^{N_n}\frac{1}{M_\nu}\left[\frac{\left[-i\hbar\nabla_\nu-\bA_\nu(\dulR,t)\right]^2}{2}+\right.\\
&\left.\left(\frac{-i\hbar\nabla_\nu\chi(\dulR,t)}{\chi(\dulR,t)}+\bA_\nu(\dulR,t)\right)\left(-i\hbar\nabla_\nu-\bA_\nu(\dulR,t)\right)\right],\nonumber
\end{align}
which acts on the parametric dependence on the nuclear positions. Since the ENCO depends on the nuclear wave function it describes \textsl{dynamical} effects of the nuclei on the electrons.
The time-dependent vector potential (TDVP),
\begin{align}\label{eqn: vector potential}
\bA_\nu(\dulR,t) = \left\langle \Phi_{\dulR}(t)\right| \left.-i\hbar\nabla_\nu\Phi_{\dulR}(t)\right\rangle_{\dulr},
\end{align}
and the time-dependent potential energy surface (TDPES),
\begin{align}\label{eqn: tdpes}
\epsilon(\dulR,t) = \left\langle\Phi_{\dulR}(t)\right|\hat H_{\mathrm{BO}}+\hat U_{\mathrm{en}}-i\hbar\partial_t\left|\Phi_{\dulR}(t)\right\rangle_{\dulr},
\end{align}
mediate the \textit{exact} coupling between the two subsystems, thus they include all effects beyond the BOA on the nuclear dynamics. In the above equations the symbol $\langle\dots\rangle_{\dulr}$ has been used to indicate integration over the electronic coordinates. The TDVP and the TDPES transform~\cite{Gross_PRL2010, *Gross_JCP2012} as standard gauge potentials, i.e.
\begin{eqnarray}
\tilde{\epsilon}(\dulR,t) &=& \epsilon(\dulR,t)+\partial_t\theta(\dulR,t)\label{eqn: transformation of epsilon} \\
\tilde{\bf A}_{\nu}(\dulR,t) &=& {\bf A}_{\nu}(\dulR,t)+\nabla_\nu\theta(\dulR,t)\,,\label{eqn: transformation of A}
\end{eqnarray}
when the electronic and nuclear wave functions transform as
\begin{equation}\label{eqn: gauge}
\begin{array}{rcl}
\chi(\dulR,t)\rightarrow\tilde\chi(\dulR,t)&=&e^{-\frac{i}{\hbar}\theta(\dulR,t)}\chi(\dulR,t) \\
\Phi_{\dulR}(\dulr,t)\rightarrow\tilde{\Phi}_{\dulR}(\dulr,t)&=&e^{\frac{i}{\hbar}\theta(\dulR,t)}
\Phi_{\dulR}(\dulr,t),
\end{array}
\end{equation}
with a gauge-like phase $\theta(\dulR,t)$ which is a real function of only nuclear positions and time. This gauge-like freedom is the only freedom in the definition of the electronic and nuclear wave functions, that can be removed by fixing the gauge.

The equations driving the evolution of the coupled electron-nuclear system within the BOA are formally similar to the exact equations~(\ref{eqn: exact eqn})~-~(\ref{eqn: tdpes}). In order to show this relation, we write the BO wave function as 
\begin{align}
\Psi^{\mathrm{BO}}(\dulr,\dulR,t) = \varphi_{\dulR}^{\mathrm{BO}}(\dulr)\chi^{\mathrm{BO}}(\dulR,t).
\end{align}
Note that $\Psi^{\mathrm{BO}}(\dulr,\dulR,t)$ is an approximation to the exact solution $\Psi(\dulr,\dulR,t)$ of the TDSE. In the BOA, the electronic wave function $\varphi_{\dulR}^{\mathrm{BO}}(\dulr)$ is an eigenstate of the BO Hamiltonian,
\begin{align}\label{eqn: BO electronic eqn}
\left[\hat H_{\mathrm{BO}}-\epsilon_{\mathrm{BO}}(\dulR)\right]\varphi_{\dulR}^{\mathrm{BO}}(\dulr) = 0,
\end{align}
with the $\dulR$-dependent eigenvalue $\epsilon_{\mathrm{BO}}(\dulR)$. The evolution of the nuclear wave function $\chi^{\mathrm{BO}}(\dulR,t)$ is given by an equation that is formally identical to Eq.~(\ref{eqn: exact nuclear eqn}),
\begin{align}\label{eqn: BO nuclear eqn}
\left[\sum_{\nu=1}^{N_n}\frac{\left[-i\hbar\nabla_\nu+\bA_\nu^{\mathrm{BO}}(\dulR)\right]^2}{2M_\nu}+\tilde\epsilon_{\mathrm{BO}}(\dulR)\right]\chi^{\mathrm{BO}}(\dulR,t)=i\hbar\partial_t\chi^{\mathrm{BO}}(\dulR,t), 
\end{align}
where a time-independent vector potential, $\bA_\nu^{\mathrm{BO}}(\dulR)$, and a time-independent scalar potential,
\begin{align}
\tilde\epsilon_{\mathrm{BO}}(\dulR)=\epsilon_{\mathrm{BO}}(\dulR)+\sum_{\nu=1}^{N_n}\frac{\hbar^2}{2M_\nu}\langle\nabla_\nu\varphi_{\dulR}^{\mathrm{BO}}|\nabla_\nu\varphi_{\dulR}^{\mathrm{BO}}\rangle_{\dulr},
\end{align}
appear. However, for a real electronic state $\varphi_{\dulR}^{\mathrm{BO}}(\dulr)$, $\bA_\nu^{\mathrm{BO}}(\dulR)$ identically vanishes, and $\tilde\epsilon_{\mathrm{BO}}(\dulR)$ contains the diagonal correction to the BO energy~\cite{Izmaylov_JCP2014, Izmaylov_JCTC2015}. These corrections to the BO energy, being $\mathcal O(M_\nu^{-1})$, are often neglected. The purpose of the following sections is to describe how equations~(\ref{eqn: BO electronic eqn}) and~(\ref{eqn: BO nuclear eqn}) are recovered from equation~(\ref{eqn: exact eqn}).

\section{Definition of the unit system}\label{sec: units}
The Hamiltonian introduced in Eq.~(\ref{eqn: hamiltonian}) is
\begin{align}\label{eqn: general hamiltonian}
\hat H =& \sum_{\nu=1}^{N_n} \frac{-\hbar^2\nabla_\nu^2}{2M_\nu} + \sum_{k=1}^{N_{el}} \frac{-\hbar^2\nabla_k^2}{2m}
-\sum_{\nu=1}^{N_n}\sum_{k=1}^{N_{el}} \alpha\hbar c\frac{Z_\nu e^2}{\left|\bR_\nu-\br_k\right|}\nonumber\\
&+\sum_{k=1}^{N_{el}}\sum_{k'>k} \alpha\hbar c\frac{e^2}{\left|\br_k-\br_{k'}\right|}
+\sum_{\nu=1}^{N_n}\sum_{\nu'>\nu} \alpha\hbar c\frac{Z_\nu Z_{\nu'}e^2}{\left|\bR_\nu-\bR_{\nu'}\right|},
\end{align}
where the first two terms on the right-hand-side are the nuclear and electronic kinetic energy operators, respectively, followed by the electron-nuclear, purely electronic and purely nuclear Coulomb interactions. As in Sec.~\ref{sec: theory}, $\nu$ and $\nu'$ label the nuclei, whereas $k$ and $k'$ label the electrons; $M_\nu$ and $m$ are the masses of the nuclei and of the electrons, respectively; the position of each nucleus is indicated as $\bR_\nu$ and that of each electron as $\br_k$; $\alpha$ is the fine structure constant, $c$ the speed of light and $Z_\nu e$ (or $Z_{\nu'}e$) is the charge corresponding to the nucleus $\nu$ (or $\nu'$).

We introduce a unit system~\cite{Ciccotti_JCP1999} where mass $M_0$, length $\lambda_0$, energy $e_0$ and charge $e$ are the fundamental units, namely
\begin{align}\label{eqn: fundamental units}
M_0=N_n^{-1}\sum_\nu M_\nu; \,\, \lambda_0=\frac{\hbar}{m\alpha c};\,\, e_0=m\alpha^2c^2; \,\, e.
\end{align}
Notice that the new units of length, energy and charge are the same used in the atomic unit (a.u.) system, whereas the average nuclear mass, rather than the electronic mass, is used as unit of mass. All other units can be derived starting from the basic ones chosen above. For instance, the units of time, action and velocity are
\begin{align}\label{eqn: derived units}
[t] = \sqrt{\frac{M_0}{e_0}}\lambda_0;\,\,[a]=\sqrt{M_0e_0}\lambda_0;\,\,[v] = \sqrt{\frac{e_0}{M_0}}.
\end{align}
In order to express Hamiltonian (\ref{eqn: general hamiltonian}) in the new units, we have to derive
the values of fundamental constants $c$ and $\hbar$ in our unit system, 
\begin{subequations} \label{eqn: c h new units}
\begin{align}
c = c'[v] = c'\sqrt{\frac{e_0}{M_0}}=c'\sqrt{\mu}\alpha c&\rightarrow c' = \frac{1}{\sqrt{\mu}\alpha} \label{eqn: c'}\\
\hbar = \hbar'[a] = \hbar'\sqrt{M_0e_0}\lambda_0  = \hbar'\frac{1}{\sqrt{\mu}}\hbar&\rightarrow \hbar' = \sqrt{\mu}.\label{eqn: h'}
\end{align}
\end{subequations}
Henceforth, primed symbols will indicate dimensionless quantities with the only exception of the electron-nuclear mass ratio $\mu = m / M_0$.
Our particular choice of units leads to two interesting results. First, the speed of light is scaled with $\mu^{-\frac{1}{2}}$, thus if $\mu\rightarrow 0$, i.e., if the nuclei become infinitely massive compared to the electrons, $c'$ tends to infinity. Second, Planck's constant is the square root of the electron-nuclear mass ratio, thus if $\mu\rightarrow 0$ also $\hbar'$ becomes zero. We will refer to the limit $\mu \to 0$ as the ``adiabatic limit''. Since $\mu$ is a dimensionless mass ratio, the adiabatic limit means that $m \ll M_0$, i.e., the mass of the electron is much smaller than the average nuclear mass.

Using Eq.~(\ref{eqn: c h new units}) it is straightforward to express the Hamitonian (\ref{eqn: general hamiltonian}) in our new unit system:
\begin{align}
\hat H = e_0&\left[\mu \sum_{\nu}-\frac{\nabla_\nu'^2}{2M_\nu'}+\sum_{k}-\frac{\nabla_k'^2}{2}-\sum_{\nu=1}^{N_n}\sum_{k=1}^{N_{el}} \frac{Z_\nu}{\left|\bR_\nu'-\br_k'\right|}\right.\nonumber \\
&+\left.\sum_{k=1}^{N_{el}}\sum_{k'>k} \frac{1}{\left|\br_k'-\br_{k'}'\right|}
+\sum_{\nu=1}^{N_n}\sum_{\nu'>\nu} \frac{Z_\nu Z_{\nu'}}{\left|\bR_\nu'-\bR_{\nu'}'\right|}\right]\nonumber\\
=e_0 &\left[\mu \hat T_n'+\hat H_{\mathrm{BO}}'\right].
\end{align}
Obviously the nuclear kinetic energy, $\mu \hat T_n'$, scales linearly with the electron-nuclear mass ratio $\mu$. This outcome is indeed intuitively expected, since we expect that in the limit of nuclei that are much heavier than the electrons the nuclear kinetic energy becomes irrelevant. We point out that Hamiltonian (\ref{eqn: general hamiltonian}) has the same form in standard atomic units. The crucial difference in using units (\ref{eqn: fundamental units}) appears when we look at the TDSE. The time derivative in our unit system reads
\begin{align}
i\hbar \partial_t = i e_0 \sqrt{\mu} \partial_{t'} ~, \label{eqn: dt new units}
\end{align}
i.e., it scales with $\sqrt{\mu}$ due to the scaling of $\hbar$. This is crucial to obtain a static electronic equation for $\mu \to 0$ when the new units are used in Eq.~(\ref{eqn: exact electronic eqn}). Similarly the spatial derivatives (momentum operators) are
\begin{align}
-i\hbar \nabla = -i \sqrt{\mu} \sqrt{M_0 e_0} \nabla' ~, \label{eqn: momentum new units}
\end{align}
in our unit system.

\section{Electronic equation}\label{sec: electronic equation}

Equipped with the results from Sec.~\ref{sec: units} we rewrite the TDVP in our new units. From Eq.~(\ref{eqn: momentum new units}) follows
\begin{align}
\mathbf A_\nu(\bR,t) \to \sqrt{\mu}\sqrt{M_0e_0}\mathbf A_\nu'(\bR',t')~,
\end{align}
with 
\begin{align}\label{eqn: vector potential new units}
\bA_\nu'(\dulR',t') = \int d\br'\,\Phi_{\bR'}'^*(\br',t)\left(-i \nabla_\nu' \Phi_{\bR'}'(\br',t')\right)~.
\end{align}
Similarly we have 
\begin{align}\label{eqn: nabla chi over chi in new units}
\frac{-i\hbar\nabla_\nu\chi(\bR,t)}{\chi(\bR,t)} \to \sqrt{\mu}\sqrt{M_0e_0}\frac{-i\nabla_\nu'\chi'(\bR',t')}{\chi'(\bR',t')},
\end{align}
which leads to the ENCO in our new units, i.e., 
\begin{align}
\hat U_{\mathrm{en}}(t) \to e_0 \mu\hat U_{\mathrm{en}}'(t')~,
\end{align}
with 
\begin{align}\label{eqn: enco in prime units}
\hat U_{\mathrm{en}}'(t') = &
\sum_{\nu=1}^{N_n}\frac{1}{M_\nu'}\left[\frac{\left[-i\nabla_\nu'-\bA_\nu'(\dulR',t')\right]^2}{2}\right.\\
&\left. + \left(\frac{-i\nabla_\nu'\chi'(\dulR',t')}{\chi'(\dulR',t')}
+ \bA_\nu'(\dulR',t')\right)\left(-i\nabla_\nu'-\bA_\nu'(\dulR',t')\right)\right].\nonumber
\end{align}
Finally, we turn to the TDPES. It is the sum of three contributions, i.e.,
\begin{align}
\epsilon(\bR,t) =& \left\langle\Phi_\bR(t)\right| \hat H_{\mathrm{BO}}+\hat U_{\mathrm{en}}-i\hbar\partial_t\left|\Phi_\bR(t)\right\rangle_\br \nonumber \\
=& \,\epsilon_{\mathrm{BO}}(\bR,t)+\epsilon_{\mathrm{bBO}}(\bR,t)+\epsilon_{\mathrm{td}}(\bR,t)~,
\end{align}
the BO term, the bBO (beyond BO) term and the td (time derivative) term, where $\epsilon_{\mathrm{bBO}}$ only contains the first term on the right-hand-side of Eq.~(\ref{eqn: enco}), since the second term does not contribute (by construction) to the expression of the TDPES. When using the new unit system, we determine the explicit dependence on $\mu$ of these three contributions, namely
\begin{align}
\epsilon(\bR,t) \to e_0\left[\epsilon_{\mathrm{BO}}'(\bR',t')+\mu\epsilon_{\mathrm{bBO}}'(\bR',t')+\sqrt{\mu}\epsilon_{\mathrm{td}}'(\bR',t')\right].
\end{align}
The electronic equation given in Eq.~(\ref{eqn: exact electronic eqn}) thus becomes
\begin{align}\label{eqn: el eqn in new units with td}
\Big(\hat H_{\mathrm{BO}}'-&\epsilon_{\mathrm{BO}}'\Big)\Phi_{\bR'}'=\left[\sqrt{\mu}\left(i\partial_{t'}+\epsilon_{\mathrm{td}}'\right)
-\mu\Big(\hat U_{\mathrm{en}}'-\epsilon_{\mathrm{bBO}}'\Big)\right]\Phi_{\bR'}'.
\end{align}
The new units single out the explicit dependence on $\mu$ of each term in the evolution equation for the electronic factor. As anticipated above, this dependence also appears in the term containing the time derivative of the electronic factor. When the adiabatic limit is approached as $\mu\rightarrow 0$, this term vanishes as the square root of the mass ratio: we recover the BOA for the coupled electron-nuclear dynamics, the electronic state is static, in the sense that the electronic system is always found in the same eigenstate (solution of the eigenvalue problem associated to $\hat H_{\mathrm{BO}}$).

The alert reader may have noticed that the ENCO [cf.\ Eq.~(\ref{eqn: enco in prime units})] contains a sum over all nuclei. Each term contains a prefactor $\mu/M'_\nu$, i.e., the ratio of the dimensionless electron mass $\mu$ and the dimensionless nuclear mass $M'_\nu$. So far we have implicitly assumed that all the $M'_\nu$ are of order one, which means that the molecule is comprised of nuclei with similar masses. If, however, the molecule contains nuclei which are much lighter than the average nuclear mass, e.g., if the molecule contains hydrogen, the corresponding $M'_\nu$ might not be of order one but instead $M'_\nu <1$. Note that this is not in conflict with our general discussion, for it simply means that the adiabatic limit may not be suited to describe the present system.

In the adiabatic limit, the time dependence is solely carried by the nuclear wave function, which evolves according to a TDSE where the potential is given by the BO energy eigenvalue at each nuclear position, as will be argued in Sec.~\ref{sec: nuclear equation}. This result depends on the viewpoint adopted in the XF, as we have already mentioned in the Introduction. The coupled equations in fact describe how the nuclei move and how the electrons follow their motion, thus the explicit time dependence of the electronic factor only arises as effect of the coupling to the nuclei. If the nuclear mass increases, then nuclear motion becomes slower and define a time scale over which the electrons are able to react instantaneously. This is the reason why the XF equations lead to the BOA in the adiabatic limit. In the time-dependent version of BOA, the electronic and nuclear systems are not treated on equal footing: the electronic equation is static, whereas the nuclear equation is an evolution equation. The electronic equation is an eigenvalue problem, satisfied only by the eigenstate of $\hat H_{\mathrm{BO}}'$ with eigenvalue $\epsilon_{\mathrm{BO}}'$ (see the left-hand-side of Eq.~(\ref{eqn: el eqn in new units with td})). Therefore, at all times during the evolution of the nuclear wave function (and also at the initial time), the electronic state must be an eigenstate of the BO Hamiltonian. Furthermore, in Eq.~(\ref{eqn: el eqn in new units with td}) the term responsible for the coupling with the nuclear motion, $\hat U_{\mathrm{en}}$, is eliminated as the mass ratio goes to zero. Clearly, non-adiabatic effects disappear in the adiabatic limit. However, if we set now the gauge such that $\epsilon_{\mathrm{td}}'=0$ (or equivalently $\epsilon_{\mathrm{td}}=0$), which is possible in general, the electronic equation becomes
\begin{align}\label{eqn: full el eqn with primes}
\Big(\hat H_{\mathrm{BO}}'-&\epsilon_{\mathrm{BO}}'\Big)\Phi_{\bR'}'=\left[\sqrt{\mu}i\partial_{t'}
-\mu\Big(\hat U_{\mathrm{en}}'-\epsilon_{\mathrm{bBO}}'\Big)\right]\Phi_{\bR'}'.
\end{align}
Non-adiabatic contributions, induced by $\hat U_{\mathrm{en}}'$, seem to appear only at a higher order than the time derivative itself. This is somehow unexpected, as the explicit dependence on time of the electronic state arises as consequence of the coupling to nuclear motion, which is encoded in the ENCO. However, it has to be borne in mind that so far we have only analyzed the explicit dependence on $\mu$ in the evolution equation for the electrons.
At this point we have only established that in the limit $\mu \to 0$ the electronic equation reduces to the static BOA for the electronic factor under the assumption that the $\mu \to 0$ limit of Eq.~(\ref{eqn: nabla chi over chi in new units}) does not diverge as $\mu^{-\frac{1}{2}}$ or stronger.
We investigate this issue in Sec.~\ref{sec: nuclear equation}.

\section{Nuclear equation}\label{sec: nuclear equation}
Let us suppose that the nuclear wave function is a coherent state, with Gaussian-shape density and propagating with a certain momentum. Then, as the adiabatic limit is approached, one would intuitively expect that the density localizes and that the motion becomes slower~\cite{Hagedorn_AM1986}. This effect can be understood classically as consequence of the increase of the nuclear mass. Therefore, in the analysis reported below, we will also take into account the dependence of the nuclear wave function on the mass ratio. We consider explicitly this dependence here, since only the nuclear term in the factorization~(\ref{eqn: factorization}) is a proper nuclear wave function; the electronic factor has simply a parametric ``conditional'' dependence on $\bR$, which we assume not being affected by the nuclear mass.
 
In order to motivate this assumption let us consider the probabilities $|\chi(\bR,t)|^2$ and $|\Phi_\bR(\br,t)|^2$, rather than the probability amplitudes. It is natural to expect that a massive nucleus yields a density which is very localized in space, associated to a large marginal probability, $|\chi(\bR,t)|^2$, in certain regions of space. As the nuclear mass decreases, the value of the probability decreases as well, consequence of the delocalization of a light particle (and of the normalization of the probability). At the same time, $|\Phi_\bR(\br,t)|^2$ is a conditional probability, providing the probability of finding the electronic configuration $\br$ at time $t$, \textsl{given that} the nuclei are in the configuration $\bR$. Such condition, translated to the language of quantum mechanics, could be expressed as a nuclear probability density in the form of a $\delta$-function centered at $\mathbf R$, as the $\delta$-function yields the certainty of finding the nuclear configuration $\mathbf R$ at time $t$. In this sense, it seems that the conditional probability can be defined without explicit knowledge of the marginal probability $|\chi(\mathbf R,t)|^2$. However, $\Phi_{\mathbf R}(\mathbf r,t)$ and $\chi(\mathbf R,t)$, and the associated probabilities, are intimately related via the Schr\"odinger equation and the fact that they yield the full wave function. Keeping this relation in mind, we will still consider that the parametric dependence of the conditional wave function is not affected by the adiabatic limit, leaving open the possibilities for further investigations in this direction.

In the same spirit as what has been presented above, we analyze the nuclear equation~(\ref{eqn: exact nuclear eqn}). To this end it is necessary to extract the dependence on $\mu$ of the quantities appearing in the equation. Remembering that we have chosen the gauge such that $\epsilon'_{\mathrm{td}}=0$ we obtain
\begin{align} \label{eqn: exact nuclear eqn new units}
\left[\sum_{\nu=1}^{N_n}\frac{\mu\left[-i\nabla_\nu'+\bA_\nu'(\dulR',t')\right]^2}{2M_\nu'}
+\epsilon_{\mathrm{BO}}'(\bR',t')+\mu\epsilon_{\mathrm{bBO}}'(\bR',t')\right]\chi'(\dulR',t')&=i\sqrt{\mu}\partial_{t'}\chi'(\dulR',t')
\end{align}
Note that Eq.\ (\ref{eqn: exact nuclear eqn new units}) is the standard TDSE with $\hbar$ replaced
by $\sqrt{\mu}$. Furthermore, the equation is linear in $\chi'(\dulR',t')$, however, it depends
on the electronic factor through the TDVP and the TDPES. In order to proceed we use the complex phase representation~\cite{Van-Vleck_PNAS1928, Gross_PRL2015, *Gross_JCTC2016, Gross_EPL2014,*Gross_JCP2014, Agostini_ADP2015} of the nuclear wave function
\begin{align}\label{eqn: complex phase representation}
\chi'(\bR',t') = \exp{\left[\frac{i}{\sqrt{\mu}}\mathcal S'(\bR',t')\right]},
\end{align}
with $\mathcal S'$ a complex function of $\bR'$ and $t'$. Inserting this expression into the nuclear TDSE~(\ref{eqn: exact nuclear eqn new units}) we arrive at
\begin{align}
-\partial_{t'}\mathcal S' = \sum_\nu\frac{1}{2M_\nu'}\Big[\left(\nabla_\nu' \mathcal S'\right)^2-i\sqrt{\mu}\nabla_\nu'^2\mathcal S'+2\mu\mathbf A_\nu'\cdot\nabla_\nu' \mathcal S'-i\mu\nabla_\nu'\cdot \mathbf A_\nu'+\mu\mathbf A_\nu'^2\Big]+\epsilon_{\mathrm{BO}}'+\mu\epsilon_{\mathrm{bBO}}'.\label{eqn: full eqn for complex phase}
\end{align}
At this point we are at the same level as in our discussion of the electronic equation in Sec.~\ref{sec: electronic equation}, i.e., we have singled out the explicit dependence on $\sqrt{\mu}$ in the evolution equation. Similarly to the discussion reported above on the adiabatic limit of the ENCO, the sum over the nuclei $\nu$ in the kinetic energy of the nuclei contains prefactors $\mu/M'_\nu$. This means that for heteronuclear molecules the validity of the adiabatic description has to be critically examined. In contrast to the evolution equation for the electronic factor we are here dealing with a proper TDSE. Accordingly we can use that the complex phase $\mathcal S'$ can be expanded~\cite{Van-Vleck_PNAS1928} as an asymptotic series in powers of $\hbar'=\sqrt{\mu}$, leading to the expression
\begin{align}\label{eqn: expansion of chi'}
\chi' = \exp{\left[\frac{i}{\sqrt{\mu}}\left(S_0'+\sqrt{\mu}S_1'+\mu S_2'+\ldots\right)\right]},
\end{align}
for the nuclear wave function. Combining equations (\ref{eqn: expansion of chi'}) and (\ref{eqn: full eqn for complex phase}) we can take the limit $\mu \to 0$ to obtain
\begin{align}\label{eqn: adiabatic limit of nuclear equation}
-\partial_{t'}S_0' = \sum_\nu\frac{1}{2M_\nu'}\left(\nabla_\nu' S_0'\right)^2 + \epsilon_{\mathrm{BO}}' ~.
\end{align}
This is formally a Hamilton-Jacobi equation for the function $S_0' $ where the potential energy is simply given by the BO contribution to the TDPES. The nuclear equation does not become a static eigenvalue equation in the adiabatic limit, as it happens for the electronic equation. The nuclear equation is still a dynamical equation which yields the adiabatic evolution of the nuclear wave function. In fact, it is easy to see that the same expression would be recovered starting from Eq.~(\ref{eqn: BO nuclear eqn}), or equivalently from
\begin{align}\label{eqn: BO nuclear eqn from XF}
\left[\sum_\nu\frac{-\mu\nabla_\nu'^2}{2M'_\nu} + \epsilon_{\mathrm{BO}}' (\dulR')\right]\chi'(\dulR',t') = i\sqrt{\mu}\partial_{t'}\chi'(\dulR',t'),
\end{align}
expressed in the new unit system, which is the nuclear TDSE~(\ref{eqn: exact nuclear eqn new units}) without the TDVP where the only potential is $\epsilon_{\mathrm{BO}}'$. In the adiabatic limit $\mu \to 0$ the nuclear TDSE turns into a Hamilton-Jacobi equation (\ref{eqn: adiabatic limit of nuclear equation}). It can be solved by means of its characteristics, corresponding to \textsl{classical} trajectories in phase space with momenta given by the gradient of $S_0'$.

Now we are able to address the issue raised at the end of Sec.~\ref{sec: electronic equation}, i.e., whether the $\mu$ dependence of the nuclear wave function spoils the conclusions regarding the $\mu \to 0$ limit of the electronic equation. Based on Eq.~(\ref{eqn: expansion of chi'}), we analyze the term in the expression of the ENCO which depends on $\chi'$. In Eq.\ (\ref{eqn: nabla chi over chi in new units}) we consider the term on the right-hand-side, removing the explicit dependence on the units of momentum, i.e., $\sqrt{M_0 e_0}$, and we insert the expansion (\ref{eqn: expansion of chi'}), to get
\begin{align}\label{eqn: expansion of nabla chi over chi}
\sqrt{\mu}\frac{-i\nabla_\nu'\chi'}{\chi'} = \nabla_\nu' S_0+\sqrt{\mu} \,\nabla_\nu' S_1+\ldots,
\end{align}
meaning that the leading term of contribution (\ref{eqn: expansion of nabla chi over chi}) is actually $\mathcal O(1)$ rather than $\mathcal O(\sqrt{\mu})$. However, this does not affect the $\mu \to 0$ limit of the electronic equation and confirms that the electronic equation reduces to the static BOA in the adiabatic limit. Furthermore, we can rewrite the expression of the ENCO as the sum of
\begin{align}\label{eqn: first term of the enco}
\sqrt{\mu} \,{\hat U_{\mathrm{en}}'^{\mathrm{I}}}= \sum_\nu \frac{\big(\nabla_\nu'S_0'\big)}{M_\nu'}\Big[\sqrt{\mu}\left(-i\nabla_\nu'-\mathbf A_\nu'\right)\Big]
\end{align}
and
\begin{align}\label{eqn: second term of the enco}
\mu \,{\hat U_{\mathrm{en}}'^{\mathrm{II}}}=\sum_\nu \frac{1}{M_\nu'}\bigg[\mu\frac{[-i\nabla_\nu'-\mathbf A_\nu']^2}{2} +\mu\Big(\big[\nabla_\nu'S_1+\mathcal O(\mu^{\frac{1}{2}})\big]+\mathbf A_\nu'\Big)\left(-i\nabla_\nu'-\mathbf A_\nu'\right)\bigg].
\end{align}
In Eq.~(\ref{eqn: full el eqn with primes}) it is clear that the ENCO appears with a prefactor $\mu$, i.e., $\mu \,\hat U'_{\mathrm{en}}$. However, due to the dependence on $\mu$ of the nuclear wave function, we can distinguish two different contributions, $\sqrt{\mu}\,{\hat U_{\mathrm{en}}'^{\mathrm{I}}}$ and $\mu\,{\hat U_{\mathrm{en}}'^{\mathrm{II}}}$, leading to the electronic equation
\begin{align}
\Big(\hat H_{\mathrm{BO}}'-&\epsilon_{\mathrm{BO}}'\Big)\Phi_{\bR'}'=\Big[\sqrt{\mu}\left(i\partial_{t'}-\hat U_{\mathrm{en}}'^{\mathrm{I}}\right)-\mu\left(\hat U_{\mathrm{en}}'^{\mathrm{II}}-\epsilon_{\mathrm{bBO}}'\right)\Big]\Phi_{\bR'}'.
\end{align}
The coupling to the nuclear motion, expressed at leading order by $\hat U_{\mathrm{en}}'^{\mathrm{I}}$, induces an explicit time dependence in the electronic equation, as described by the first term on the right-hand-side of the above equation (the term $\mathcal O(\sqrt{\mu})$). Note that this resolves also the puzzle mentioned earlier in that it shows that the ENCO enters already at order $\sqrt{\mu}$. It is worth stressing again that in Eqs.~(\ref{eqn: first term of the enco}) and~~(\ref{eqn: second term of the enco}) the ratios $\sqrt{\mu} / M'_\nu$ and $\mu / M'_\nu$ enter, respectively~\footnote{Note that since $\mu$ and $M'_\nu$ are dimensionless quantities the ratio $\sqrt{\mu} / M'_\nu$ is meaningful.}. As mentioned earlier this is important only if some of the nuclei in the molecule are much lighter than others.

It is clear that the leading order correction to the adiabatic limit is $\mathcal O(\sqrt{\mu})$ and only affects the dynamics of the electronic system. The contribution to the energy~\cite{Scherrer_JCP2015, Scherrer_TBS2016} of the term $\mathcal O(\sqrt{\mu})$ is identically zero, since
\begin{align}\label{eqn: td terms to zero}
\left\langle \Phi'_{\mathbf R'}(t')\left|i\partial_{t'}-\hat U_{\mathrm{en}}'^{\mathrm{I}}\right|\Phi'_{\mathbf R'}(t')\right\rangle_{\mathbf r'}=0.
\end{align}
The first term on the left-hand-side of Eq.~(\ref{eqn: td terms to zero}) is zero due to the choice of the gauge, the second is zero by construction. The correction to the BO energy, namely $\epsilon_{\mathrm{bBO}}(\bR,t)$, only appears at a higher order, i.e. $\mathcal O(\mu)$. 

In the adiabatic limit $\mu\rightarrow 0$, $\epsilon_{\mathrm{BO}}'$ is determined as the solution of an eigenvalue problem with $\hat H_{\mathrm{BO}}'$. This PES is not a time-dependent quantity and appears in the nuclear equation~(\ref{eqn: adiabatic limit of nuclear equation}) as unique effect due to the electrons. In fact, as we have shown in Eq.~(\ref{eqn: full eqn for complex phase}), the TDVP appears at $\mathcal O(\mu)$.

\section{Numerical analysis}\label{sec: results}
We have performed numerical simulations with the aim of analyzing the dependence of the nuclear wave function on the mass ratio $\mu$. In particular we focus here on the term that in the ENCO depends explicitly on the nuclear wave function. As clear from Eq.~(\ref{eqn: expansion of nabla chi over chi}), this contribution,
\begin{align}\label{eqn: nuclear momentum}
\frac{-i\nabla_\nu'\chi'}{\chi'} = \mu^{-\frac{1}{2}}\Big[\nabla_\nu'S_0'+\mathcal O(\sqrt{\mu})\Big],
\end{align}
is at least $\mathcal O(\mu^{-\frac{1}{2}})$, and it is the leading non-adiabatic, or beyond BO, effect on electronic dynamics. 

In Eq.~(\ref{eqn: adiabatic limit of nuclear equation}) we have shown that $S_0'$ satisfies a Hamilton-Jacobi equation, where $\nabla_\nu'S_0'$ appears as momentum contribution. In previous work~\cite{Gross_EPL2014, *Gross_JCP2014, Gross_PRL2015, *Gross_JCTC2016} based on the $\hbar$-expansion of the complex phase representation of the nuclear wave function, in fact, we have shown that the lowest order term can be interpreted as the classical action, whose gradient is indeed the momentum evaluated along the classical trajectory. In the present work, we have related the $\hbar$-expansion to the $\sqrt{\mu}$-expansion. It seems therefore natural to interpret also in this context the leading contribution $\nabla_\nu' S_0'$, in Eq.~(\ref{eqn: nuclear momentum}), as the classical nuclear momentum expressed in the unit system defined in Eq.~(\ref{eqn: fundamental units}). The dependence on the electron-nuclear mass ratio is shown analytically to be $\mu^{-\frac{1}{2}}$. Several numerical schemes have been proposed~\cite{Tully_JCP1990, Marx_PRL2002, Tavernelli_PRL2002, Ehrenfest_ZP1927, Nafie_JCP1983, Scherrer_JCTC2013, tully, Ciccotti_JCP2002, Gross_EPL2014,*Gross_JCP2014, Scherrer_JCP2015, Schild_JPCA2016} which consider only non-adiabatic effects induced by the dependence of electronic motion on the nuclear momentum. Higher order terms~\cite{Gross_PRL2015, *Gross_JCTC2016} are usually discarded. Highlighting the dependence on the electron-nuclear mass ratio of all terms beyond BO in the framework of the XF has led us to rationalize the importance of the nuclear momentum in inducing non-adiabatic effects.

\subsection{Simulation details}\label{sec: details}
\begin{figure}
 \begin{center}
 \includegraphics*[width=0.5\textwidth]{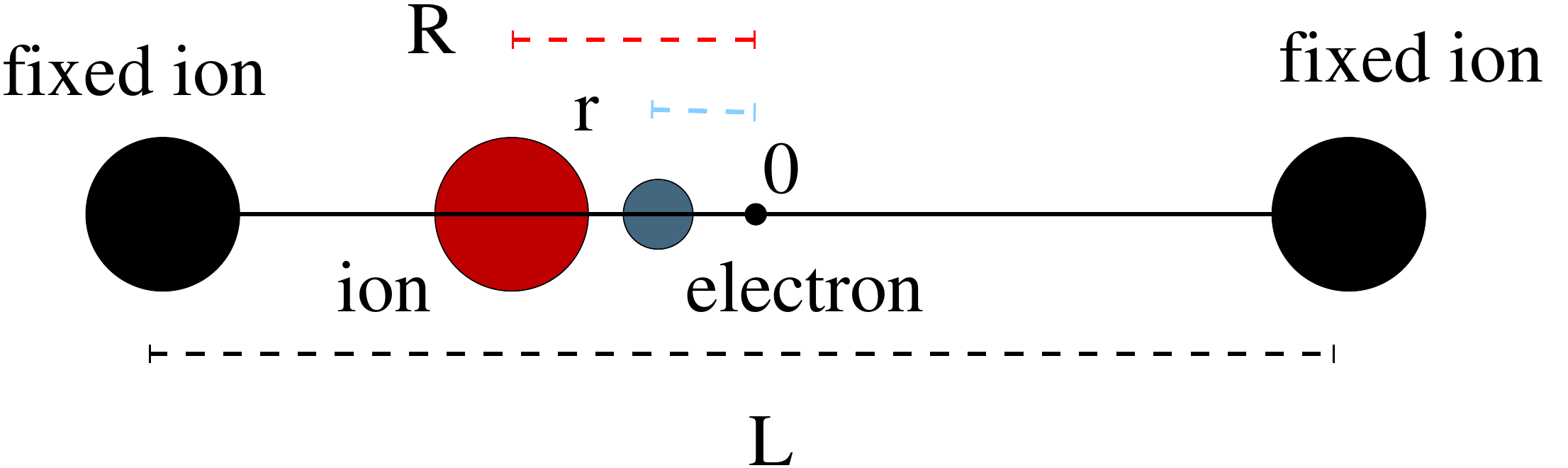}
 \caption{Schematic representation of the model proton-coupled electron transfer.}
 \label{fig: model}
 \end{center}
\end{figure}

In order to numerically validate the discussion presented in Sec.~\ref{sec: nuclear equation}, we have simulated the non-adiabatic dynamics of a simple system representing a model for the process of proton-coupled electron transfer~\cite{MM}. The system is composed by three ions and one electron in one dimension, as shown in Fig.~\ref{fig: model}: two ions are fixed at a distance $L=19.0$~a$_0$, the moving ion interacts with the fixed ions via a Coulomb potential, the moving electron interacts with all ions via soft-Coulomb potentials. The Hamiltonian of the system reads
\begin{align}
 \hat{H}(r,R)= &-\frac{\mu}{2}\frac{\partial^2}{\partial R^2}-\frac{1}{2}\frac{\partial^2}{\partial r^2}  +
 \frac{1}{\left|\frac{L}{2}-R\right|}+\frac{1}{\left|\frac{L}{2} + R\right|}\nonumber\\
& -\frac{\mathrm{erf}\left(\frac{\left|R-r\right|}{R_c}\right)}{\left|R - r\right|}
 -\frac{\mathrm{erf}\left(\frac{\left|r-\frac{L}{2}\right|}{R_r}\right)}{\left|r-\frac{L}{2}\right|}
 -\frac{\mathrm{erf}\left(\frac{\left|r+\frac{L}{2}\right|}{R_l}\right)}{\left|r+\frac{L}{2}\right|}.\label{eqn: metiu-hamiltonian}
\end{align}
The parameters of the soft-Coulomb interaction, $R_c$ (relative to the moving ion), $R_r$ (relative to the right ion) and $R_l$ (relative to the left ion), are chosen in order to induce non-adiabatic coupling mainly between the two lowest electronic states. The values of the parameters are $R_r=R_l=3.5$~a$_0$, whereas, in order to investigate different situations of coupling, $R_c=4.0, 5.5, 7.0$~a$_0$. We refer to the three situations as ``weak'', ``intermediate'' and ``strong'' coupling regimes, respectively. We will prove numerically that the theoretical development discussed in the present work applies to all situations.

\begin{figure}
\centering
\includegraphics[width=.7\textwidth,angle=270]{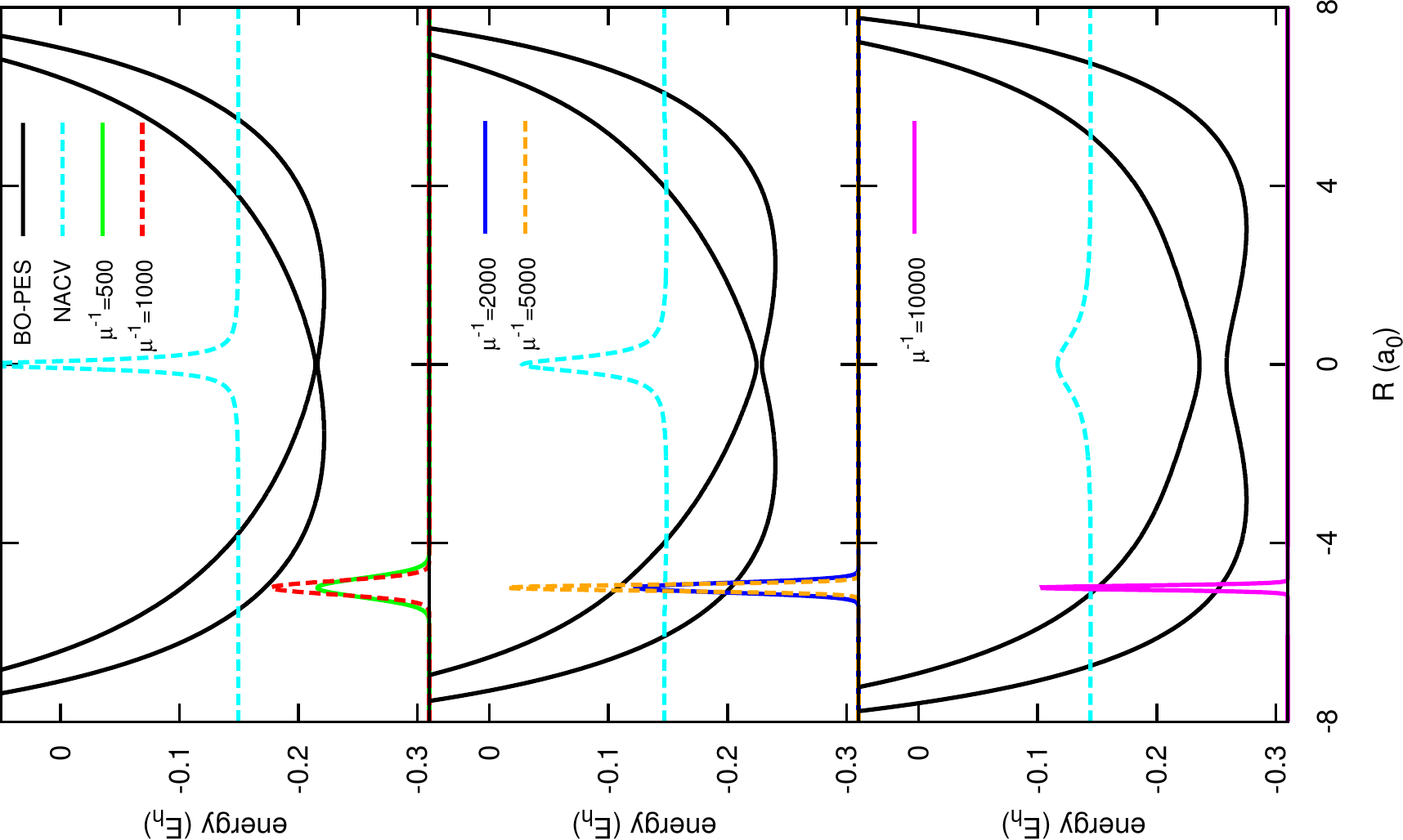}
\caption{Adiabatic potential energy surfaces corresponding to the two lowest electronic states (black lines), non-adiabatic coupling vectors (reduced 20 times) between the two lowest electronic states (cyan lines) for the strong coupling (top), intermediate coupling (middle) and weak coupling (bottom) situations; initial nuclear densities for all values of $\mu^{-1}=$ 500 (green line, reduced 20 times), 1000 (red line, reduced 20 times), 2000 (blue line, reduced 20 times), 5000 (orange line, reduced 20 times), 10000 (magenta line, reduced 40 times). The curves corresponding to the coupling vectors and to the densities have been shifted along the $y$-axis to superimpose them to the energy curves. The energy is measured in Hartree ($E_h$).}
\label{fig: bo}
\end{figure}

The BO PESs, along with the non-adiabatic coupling vector (NACV) $\langle\varphi_R^{(0)}|\partial_R\varphi_R^{(1)}\rangle_r$ between the ground $\varphi_R^{(0)}(r)$ and first excited $\varphi_R^{(1)}(r)$ (real) adiabatic states, are shown in Fig.~\ref{fig: bo} for the three coupling situations.
Dynamics is simulated by solving the full TDSE, using as initial condition $\Psi(r,R,t=0)=G_\sigma(R-R_0)\varphi_R^{(0)}(r)$, where $G_\sigma(R-R_0)$ is a real normalized Gaussian centered at $R_0=$ 5.0~a$_0$ with variance $\sigma$. Different values of the nuclear mass have been chosen, such that $\mu^{-1}=$ 500, 1000, 2000, 5000, 10000.

The simulations are performed by fixing the energy of the systems with different mass ratios such that all differences in the dynamics can be solely attributed to the difference in $\mu$. With this idea in mind, we have constructed a set of initial conditions with the same total energy for all values of $\mu$. The expression of the total energy of the electron-nuclear system can be written as
\begin{align}
E(t=0)=\left\langle \Psi(t=0)\right|\hat H\left|\Psi(t=0)\right\rangle_{r,R}=\int dR \,G_\sigma(R-R_0)\left[\hat T_n+\epsilon_{\mathrm{BO}}^{(0)}(R)\right]G_\sigma(R-R_0).
\end{align}
The Gaussian wave packets are real, thus yielding an average zero momentum. The contributions to the kinetic energy due to the spread of the Gaussian, i.e., $\mu/(4\sigma^2)$ expressed in the unit system~(\ref{eqn: fundamental units}), for different $\mu$ are determined as follows: given $\sigma_0=$ 0.15~a$_0$ (this is an arbitrary choice, made simply for numerical convenience) for $\mu_0^{-1}=$ 2000, all other values of the variance are selected according to $\sigma=\sigma_0\sqrt{\mu/\mu_0}$. It is then guaranteed that $\mu/(4\sigma^2)$ is constant for all $\mu$. The initial nuclear densities are shown in Fig.~\ref{fig: bo} for all values of the mass ratio and are all the same in the three cases studied here. The part concerning the potential energy is more complicated, because quantum mechanically it depends on the particular shape of the ground state potential and on the spreading of the nuclear density. However, the more the nuclear density localizes at $R_0$, the more the potential energy only depends on $R_0$. Therefore, the total (conserved) energy of the system results to be $E \approx -0.166$ for the strong coupling case, $E \in (-0.196,-0.197)$ for the intermediate coupling case, and $E\in(-0.246,-0.247)$ for the weak coupling case, for all values of the mass ratio.

\begin{figure}
\centering
\includegraphics[width=.7\textwidth,angle=270]{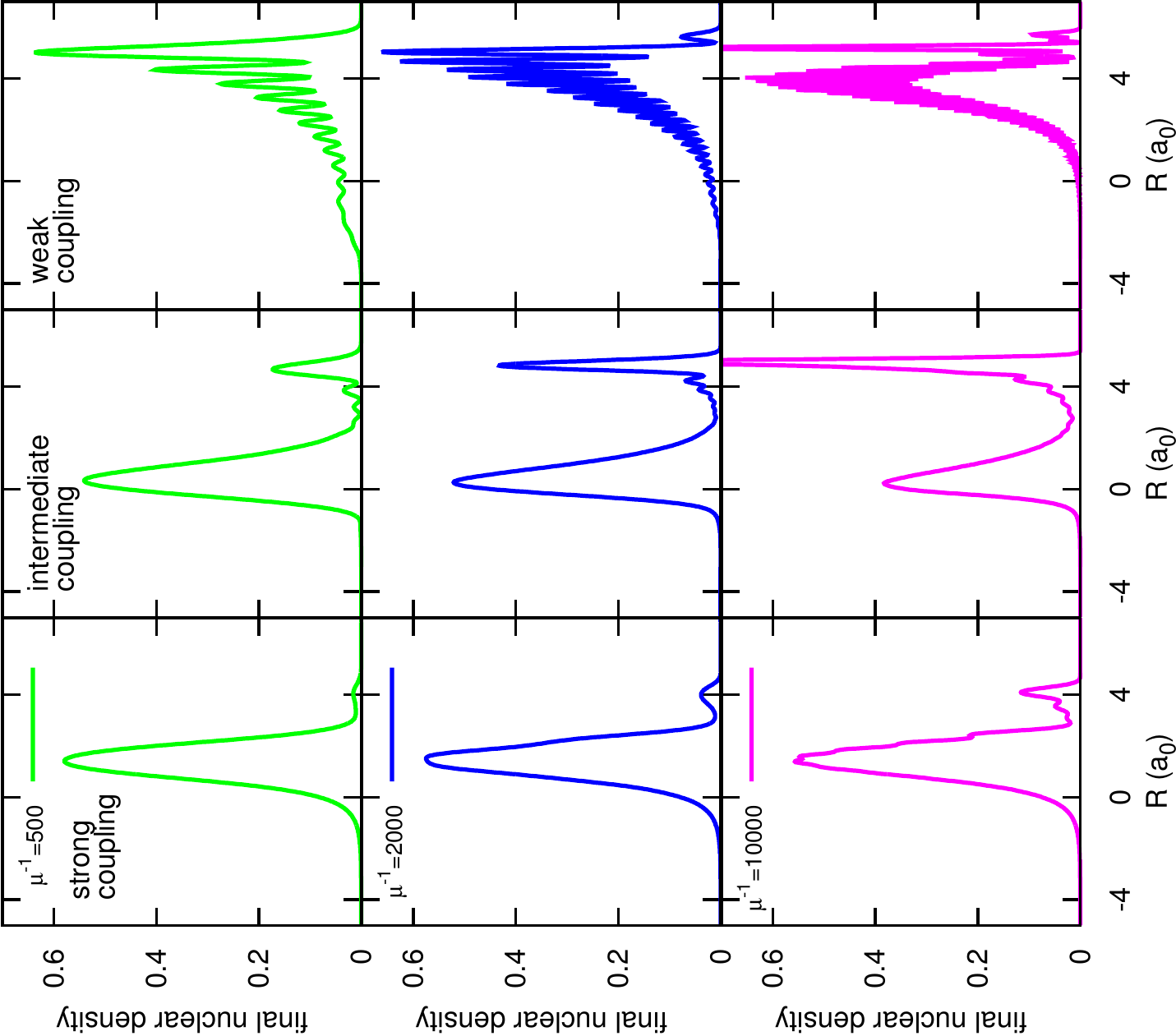}
\caption{Nuclear densities in all coupling situations for $\mu^{-1}=$ 500 (green lines), 2000 (blue lines) and 10000 (magenta lines) at the final time-step. The figure shows that the densities are indeed different for different values of the mass ratio but they are all located in the same region. This follows from the fact that the integration time-steps have been scaled with the mass ratio as discussed in the text.}
\label{fig: final steps}
\end{figure}

The split-operator-technique~\cite{spo} is used to solve the TDSE for $\Psi(r,R,t)$. The length of the simulation is chosen such that the nuclear wave packet passes through the avoided crossing at the origin only once, until the non-adiabatic process is complete. The final values of the populations of the ground state are reported in Table~\ref{tab: populations} for all coupling strengths and for all values of $\mu^{-1}$. As $\mu^{-1}$ increases the non-adiabatic event is less effective as the nuclear wave packet tends to move adiabatically on the ground state potential rather than transfer onto the excited state.
\begin{table}
\begin{tabular}{|c||c|c|c|}
\hline
$\mu^{-1}$ & strong & intermediate & weak \\
\hline
\hline
500     & 0.02 & 0.13 & 1.0 \\
1000   & 0.03 & 0.18 & 1.0 \\
2000   & 0.04 & 0.24 & 1.0 \\
5000   & 0.07 & 0.35 & 1.0 \\
10000 & 0.10 & 0.45 & 1.0 \\
\hline
\end{tabular}
\caption{Final populations of the electronic ground state for all values of $\mu^{-1}$ and for the different non-adiabatic coupling strengths.}
\label{tab: populations}
\end{table}
In order to reach the same final position in the same number of integration steps, the time-steps have been selected to satisfy the relation $dt=dt_0\sqrt{\mu_0/\mu}$, with $\mu_0=2000$ and $dt_0=$0.024~fs (0.1~a.u.), similarly to what has been done for the variance of the initial nuclear densities. This rule follows from the dependence of the time variable on the mass ratio, as shown in Eq.~(\ref{eqn: dt new units}). We obtain that as the nuclear mass increases also the time-step, and therefore the length of the simulation, increases. Figure~\ref{fig: final steps} shows for $\mu^{-1}=$ 500, 2000, 10000 that indeed the nuclear densities reach the same positions after the same number of integration time-steps.

\subsection{Dependence on the mass ratio}
In the situations described above, we have computed the average nuclear momentum for the different values of the electron-nuclear mass ratio. The leading term that in the expression of the ENCO induces non-adiabatic effects on electronic dynamics, i.e. $-i\hbar\nabla_\nu\chi/\chi$, can be interpreted classically as the nuclear momentum. The order of this leading term has been shown in Eq.~(\ref{eqn: nuclear momentum}) to be $\mu^{-\frac{1}{2}}$, when the unit system defined in Sec.~\ref{sec: units} is introduced. Therefore, we have identified an observable, the average nuclear momentum, to  be used as an indicator of the validity of this observation. We expect that the nuclear momentum is a linear function of $\mu^{-\frac{1}{2}}$.

\begin{figure}
\centering
\includegraphics[width=.7\textwidth,angle=270]{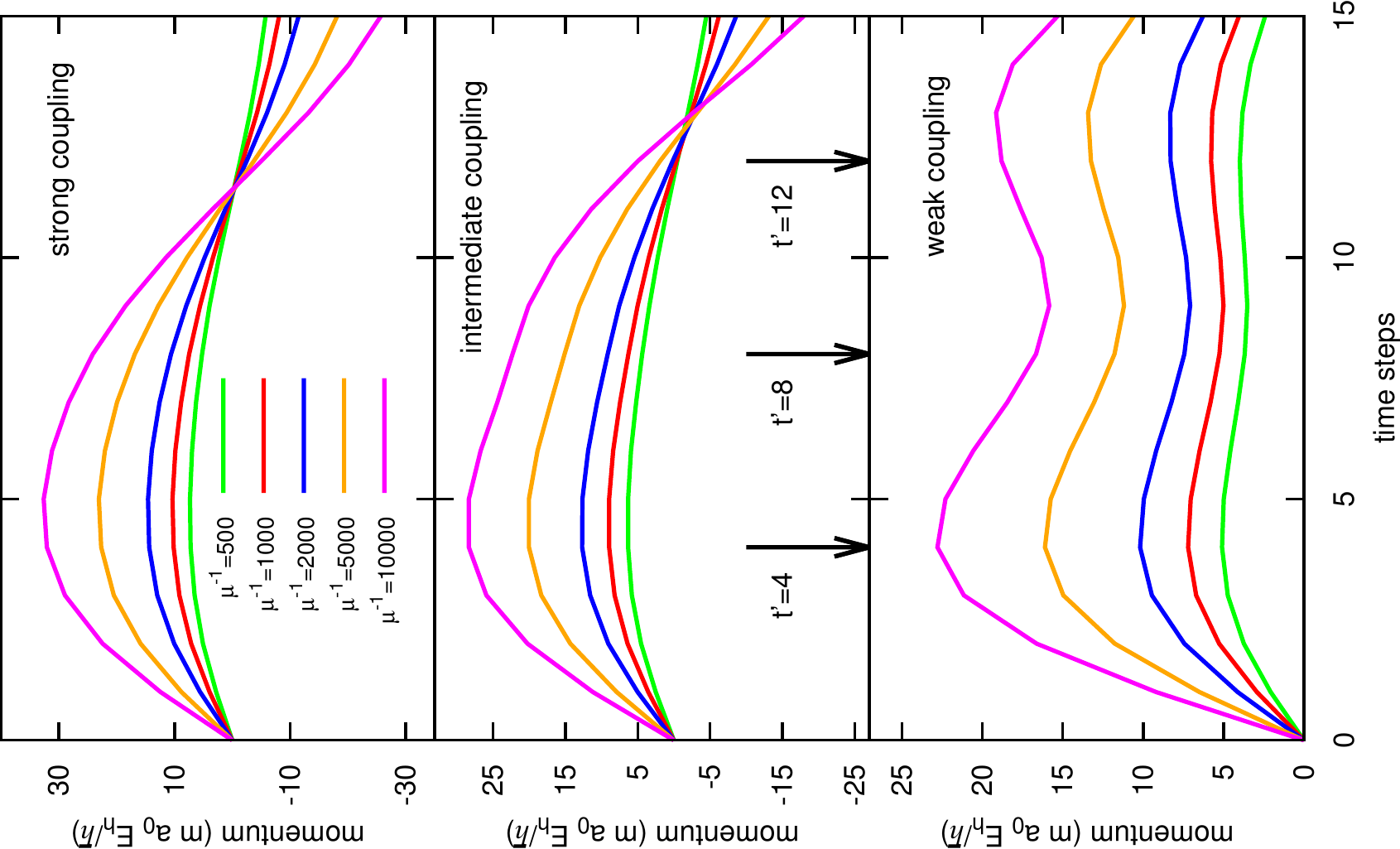}
\caption{Average nuclear momentum as function of the integration time-steps for the different coupling situations and for all values of the mass ratio. The color code for different values of $\mu^{-1}$ is the same as in Fig.~\ref{fig: bo}. The arrows in the middle panel, labeled by $t'=4$, $t'=8$ and $t'=12$, indicate the time-steps that have been selected for the analysis shown in Fig.~\ref{fig: mu-dependence}.}
\label{fig: momentum}
\end{figure}

First of all, let us show the average nuclear momentum computed along the dynamics in the three non-adiabatic situations and for all values of $\mu^{-1}$. The momentum is plotted as function of the integration time-steps. We observe in Fig.~\ref{fig: momentum} that, according to the discussion presented in Sec.~\ref{sec: details}, the behavior of the nuclear momentum is the same for all values of the mass ratio, but indeed the absolute values are different. It is such difference that we will investigate further, because it can be uniquely ascribed to the difference in the nuclear masses.
\begin{figure}
\centering
\includegraphics[width=.7\textwidth,angle=270]{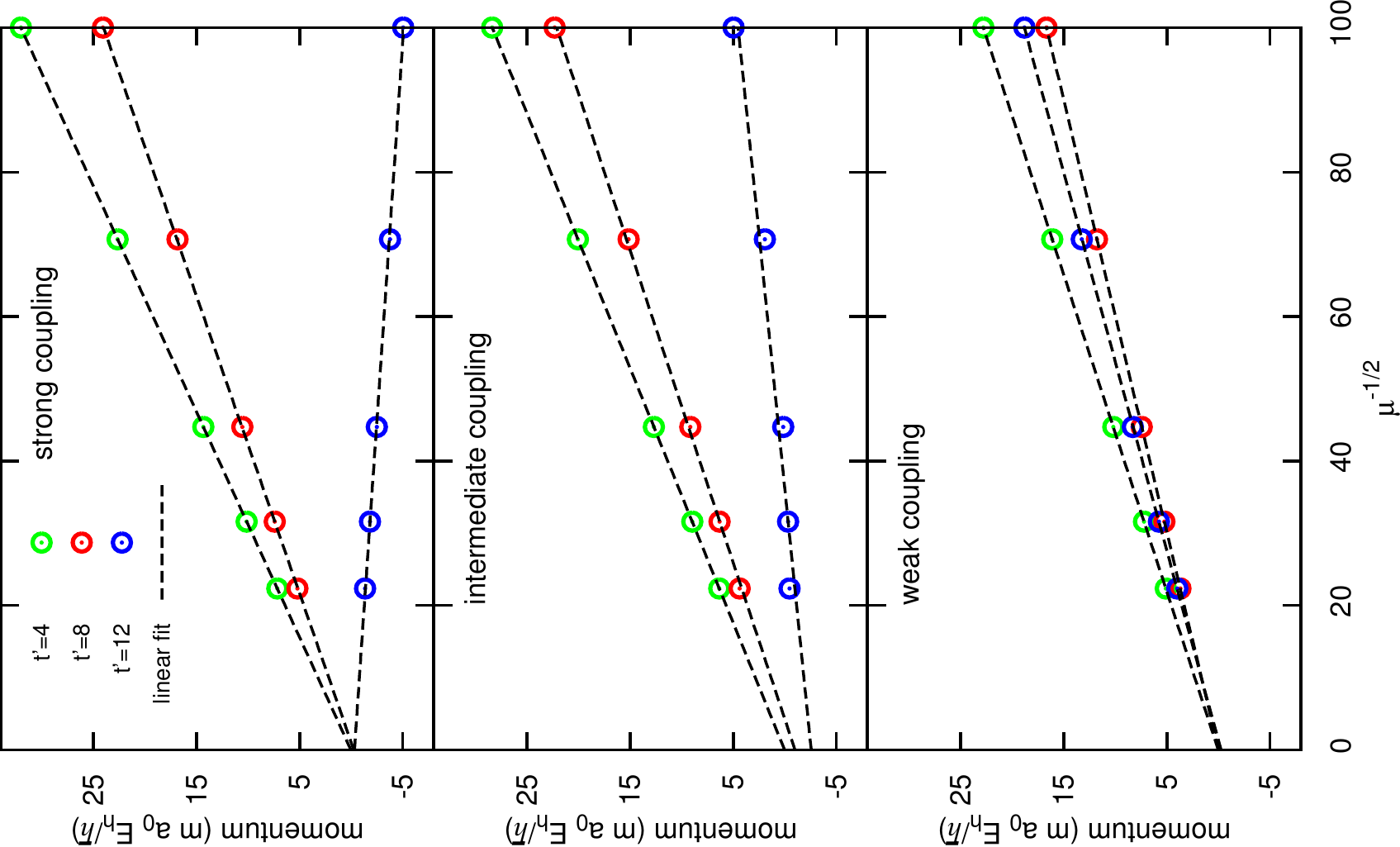}
\caption{At the time-steps selected in Fig.~\ref{fig: momentum}, namely $t'=4$ (green dots), $t'=8$ (red dots) and $t'=12$ (blue dots), the values of the average nuclear momentum is plotted as function of $\mu^{-\frac{1}{2}}$. The dashed black lines are the results of the linear fit.}
\label{fig: mu-dependence}
\end{figure}
To this end, we select three time-steps, that are indicated by the arrows in the middle panel of Fig.~\ref{fig: momentum}. At those time-steps, we plot the values of the momentum as function of $\mu^{-\frac{1}{2}}$ in Fig.~\ref{fig: mu-dependence}.
The dependence on the inverse of the square root of the mass ratio is clearly linear, as predicted by the analysis discussed in this work.

\section{Conclusions}\label{sec: conclusions}
We have developed a procedure that shows how the BOA is recovered from the XF in the limit of infinite nuclear mass, $\mu\rightarrow0$, the adiabatic limit. The adiabatic hypothesis is based on the assumption that the coupled electron-nuclear motion can be split by solving first the electronic problem at a fixed nuclear geometry, and then the nuclear problem where the potential is the eigenvalue of the electronic Hamiltonian for that geometry.

The XF provides equations of motion for the electronic and nuclear wave functions, thus we have been able to analyze separately the dependence of the two equations on $\mu$ and to show that indeed all terms inducing an explicit coupling between electronic and nuclear motion disappear in the adiabatic limit. Therefore, we have formally and in general shown what has been already observed numerically~\cite{Min_PRL2014} in a particular case, namely that the XF contains features of the BOA in the infinite nuclear mass limit. 

Non-adiabatic, or beyond BO, effects appear at different orders in $\sqrt\mu$. The leading terms only affect electronic dynamics, but do not contribute to the potential that is used in the nuclear equation. Such terms depend explicitly on the nuclear wave function, and the dependence can be approximated as the (classical) nuclear momentum. Therefore, we have justified in a rigorous way that in the limit of small, but finite, $\mu$, non-adiabatic effects can be treated perturbatively, an idea that has been already employed to correct vibrational spectra~\cite{Scherrer_JCP2015, Scherrer_TBS2016} and to compute electronic current densities~\cite{Schild_JPCA2016} beyond the BOA. In these cases, only the lowest order term, i.e., $\mathcal O(\sqrt{\mu})$, in the electronic equation has been considered, exactly in the form derived here, while the nuclear problem is treated within the adiabatic approximation. The perturbed electronic Hamiltonian induces in this way corrections to electronic wave function that are purely imaginary. This property is essential to obtain an electronic state sustaining a current~\cite{Schild_JPCA2016} and yielding a non-vanishing TDVP~\cite{Scherrer_JCP2015}. In turn, the TDVP has been expressed~\cite{Scherrer_TBS2016} as an electronic mass contribution in the nuclear Hamiltonian, up to within the lowest order in the perturbation. Furthermore, we have observed that the coefficients appearing in the expansion in powers of $\sqrt{\mu}$ depend on the nuclear masses. As those nuclear masses, $M'_{\nu}$, are unit-less quantities, they can be used to estimate the importance of the corresponding nuclei for such coefficients, and thereby partition the beyond BO effects among different nuclei. This observation deserves further investigation as it may prove useful to quantify the role of light nuclei in non-adiabatic corrections.

We have presented an analytical derivation of the adiabatic limit and supported some of the observations with a numerical analysis. In particular, we have tested the dependence on $\sqrt{\mu}$ of the leading order correction in the electronic equation, related to the nuclear momentum.

In conclusion we have established that the framework of the exact factorization of the electron-nuclear wave function reduces to the static Born-Oppenheimer approximation for the electronic wave function in the limit of vanishing electron-nuclear mass ratio. Moreover, we have shown that in the same limit the nuclear equation reduces to the classical limit of the time-dependent Sch\"odinger equation with the Born-Oppenheimer potential energy surface entering as potential for the classical nuclei. Going beyond the adiabatic limit in both equations the leading order corrections enter with the square root of the electron-nuclear mass ratio.

\section*{Acknowledgements}
The authors gratefully acknowledge discussions with Hardy Gross and Rodolphe Vuilleumier.

\end{document}